\documentclass[astrobib,amssymb,usenatbib]{mn2e}
\usepackage{times} 
\usepackage{latexsym,amsmath,amssymb,amsfonts} 
\usepackage{graphicx} 
\usepackage{fancyhdr}  
\usepackage{natbib}  
\usepackage{supertabular}
\usepackage{longtable}
\usepackage{url}
\usepackage{dcolumn}
\usepackage{morefloats}
\usepackage{hyperref}  
\begin{document}
\newcommand{\hi}{\mbox{\tiny H\,{\sc i}}}
\newcommand{\HI}{\mbox{H\,{\sc i}}}
\newcommand{\HII}{\mbox{H\,{\sc ii}}}  
\newcommand{\Htot}{\mbox{{\rm H}}}
\newcommand{\HeI}{\mbox{He\,{\sc i}}}
\newcommand{\HeII}{\mbox{He\,{\sc ii}}}
\newcommand{\HeIII}{\mbox{He\,{\sc iii}}}  
\newcommand{\OI}{\mbox{O\,{\sc i}}}
\newcommand{\OII}{\mbox{O\,{\sc ii}}}
\newcommand{\OIII}{\mbox{O\,{\sc iii}}}
\newcommand{\OIV}{\mbox{O\,{\sc iv}}}
\newcommand{\OV}{\mbox{O\,{\sc v}}}
\newcommand{\OVI}{\mbox{O\,{\sc vi}}}
\newcommand{\OVII}{\mbox{O\,{\sc vii}}}
\newcommand{\OVIII}{\mbox{O\,{\sc viii}}} 
\newcommand{\CI}{\mbox{C\,{\sc i}}}
\newcommand{\CII}{\mbox{C\,{\sc ii}}}
\newcommand{\CIII}{\mbox{C\,{\sc iii}}}
\newcommand{\CIV}{\mbox{C\,{\sc iv}}}
\newcommand{\CV}{\mbox{C\,{\sc v}}}
\newcommand{\CVI}{\mbox{C\,{\sc vi}}}  
\newcommand{\SiII}{\mbox{Si\,{\sc ii}}}
\newcommand{\SiIII}{\mbox{Si\,{\sc iii}}}
\newcommand{\SiIV}{\mbox{Si\,{\sc iv}}}
\newcommand{\SiXII}{\mbox{Si\,{\sc xii}}}   
\newcommand{\SII}{\mbox{S\,{\sc ii}}}
\newcommand{\SIII}{\mbox{S\,{\sc iii}}}
\newcommand{\SIV}{\mbox{S\,{\sc iv}}}
\newcommand{\SV}{\mbox{S\,{\sc v}}}
\newcommand{\SVI}{\mbox{S\,{\sc vi}}}  
\newcommand{\NI}{\mbox{N\,{\sc i}}}   
\newcommand{\NII}{\mbox{N\,{\sc ii}}}   
\newcommand{\NIII}{\mbox{N\,{\sc iii}}}     
\newcommand{\NIV}{\mbox{N\,{\sc iv}}}   
\newcommand{\NV}{\mbox{N\,{\sc v}}}    
\newcommand{\PV}{\mbox{P\,{\sc v}}} 
\newcommand{\NeIV}{\mbox{Ne\,{\sc iv}}}   
\newcommand{\NeV}{\mbox{Ne\,{\sc v}}}   
\newcommand{\NeVI}{\mbox{Ne\,{\sc vi}}}   
\newcommand{\NeVII}{\mbox{Ne\,{\sc vii}}}   
\newcommand{\NeVIII}{\mbox{Ne\,{\sc viii}}}   
\newcommand{\neviii}{\mbox{\tiny Ne\,{\sc viii}}}   
\newcommand{\NeIX}{\mbox{Ne\,{\sc ix}}}   
\newcommand{\NeX}{\mbox{Ne\,{\sc x}}} 
\newcommand{\MgI}{\mbox{Mg\,{\sc i}}}
\newcommand{\MgII}{\mbox{Mg\,{\sc ii}}}  
\newcommand{\MgX}{\mbox{Mg\,{\sc x}}}   
\newcommand{\FeII}{\mbox{Fe\,{\sc ii}}}  
\newcommand{\FeIII}{\mbox{Fe\,{\sc iii}}}   
\newcommand{\NaIX}{\mbox{Na\,{\sc ix}}}   
\newcommand{\ArVIII}{\mbox{Ar\,{\sc viii}}}   
\newcommand{\AlXI}{\mbox{Al\,{\sc xi}}}   
\newcommand{\CaII}{\mbox{Ca\,{\sc ii}}}  
\newcommand{\zabs}{$z_{\rm abs}$}
\newcommand{\zmin}{$z_{\rm min}$}
\newcommand{\zmax}{$z_{\rm max}$}
\newcommand{\zqso}{$z_{\rm QSO}$}
\newcommand{\degree}{\ensuremath{^\circ}}
\newcommand{\lapp}{\mbox{\raisebox{-0.3em}{$\stackrel{\textstyle <}{\sim}$}}}
\newcommand{\gapp}{\mbox{\raisebox{-0.3em}{$\stackrel{\textstyle >}{\sim}$}}}
\newcommand{\be}{\begin{equation}}
\newcommand{\en}{\end{equation}}
\newcommand{\di}{\displaystyle}
\def\tworule{\noalign{\medskip\hrule\smallskip\hrule\medskip}} 
\def\onerule{\noalign{\medskip\hrule\medskip}} 
\def\bl{\par\vskip 12pt\noindent}
\def\bll{\par\vskip 24pt\noindent}
\def\blll{\par\vskip 36pt\noindent}
\def\rot{\mathop{\rm rot}\nolimits}
\def\alf{$\alpha$}
\def\lam{$\lambda$}
\def\refff{\leftskip20pt\parindent-20pt\parskip4pt} 
\newcommand{\sqcm}{cm$^{-2}$}  
\newcommand{\lya}{Ly$\alpha$}
\newcommand{\lyb}{Ly$\beta$}
\newcommand{\lyg}{Ly$\gamma$}
\newcommand{\lyd}{Ly$\delta$} 
\def\kms{km~s$^{-1}$}
\def\zem{$z_{\rm em}$}
\def\vrel{$v_{\rm rel}$}
\def\cmsq{cm$^{-2}$}
\def\cmcb{cm$^{-3}$}
\def\etal{et~al.\ }
\newcommand{\CLOUDY}{\mbox{\scriptsize{CLOUDY}}}
\title[]{$HST/$COS detection of a \NeVIII\ absorber towards PG~1407+265:
An unambiguous tracer of collisionally ionized hot gas?
\thanks{Based on observations made with the NASA/ESA {\sl Hubble Space Telescope}, 
obtained from the data archive at the Space Telescope Science Institute, which is operated by the Association of Universities for Research
in Astronomy, Inc., under NASA contract NAS 5-26555.}}   
\author[Hussain et al.]
{
\parbox{\textwidth}{ 
T. Hussain$^{1}$,  
S. Muzahid$^{2}$,
A. Narayanan$^{3}$,
R. Srianand$^{4}$,
B. P. Wakker$^{5}$,  
J. C. Charlton$^{2}$, and    
A. Pathak$^{1}$    
} 
\vspace*{10pt}\\ 
$^{1}$Department of Physics, Tezpur University, Tezpur 784\,028, India  \\  
$^{2}$The Pennsylvania State University, 525 Davey Lab, University Park, 
State College, PA 16802, USA \\  
$^{3}$Indian Institute of Space Science \& Technology, Thiruvananthapuram 695\,547, India\\
$^{4}$ Inter-University Centre for Astronomy and Astrophysics, Post Bag 4, 
Ganeshkhind, Pune 411\,007, India \\ 
$^{5}$Department of Astronomy, University of Wisconsin-Madison, 475 North Charter St., 
Madison, WI 53706, USA \\
}   
\date{Accepted. Received; in original form } 
\maketitle
\label{firstpage}


\begin{abstract}  
 
We report the detection of \NeVIII\ in a \zabs\ = 0.59961 absorber towards the QSO PG1407+265 (\zem = 0.94). 
Besides \NeVIII, absorption from \HI\ Lyman series lines (\HI~\lam1025 -- \lam915), several other low (\CII, \NII, \OII\ and \SII), 
intermediate (\CIII, \NIII, \NIV, \OIII, \SIV\ and \SV) and high (\SVI, \OVI\ and \NeVIII) ionization metal lines are detected.
Disparity in the absorption line kinematics between different ions implies that the absorbing gas comprises of multiple ionization phases.
The low and the intermediate ions (except \SV) trace a 
compact ($\sim$~410 pc), metal-rich ($Z \sim Z_{\odot}$) and over-dense ($\log n_{\rm H} \sim -2.6$) photoionized region that sustained star-formation
for a prolonged period. 
The high ions, \NeVIII\ and \OVI,  can be explained as arising in a low density ($-5.3 \leqslant \log n_{\rm H} \leqslant -5.0$), 
metal-rich ($Z\gtrsim Z_{\odot}$) and diffuse ($\sim$~180 kpc) photoionized gas. The \SV, \SVI\ and \CIV\ (detected in the FOS spectrum) require
an intermediate photoionization phase with $-4.2 < \log n_{\rm H} < -3.5$. 
Alternatively, a pure collisional ionization model, as used to explain the previous known \NeVIII\ absorbers, with $5.65< \log T <5.72$,  
can reproduce the \SVI, \OVI\ and \NeVIII\ column densities simultaneously in a single phase.
However, even such models require an intermediate phase to reproduce any observable \SV\ and/or \CIV. 
Therefore, we conclude that when multiple phases are present, the presence of \NeVIII\ is not necessarily an 
unambiguous indication  of collisionally ionized hot gas.\\
\end{abstract}   

\begin{keywords}  
galaxies:formation -- galaxies:halos -- quasars:absorption lines 
-- quasar:individual (PG~1407+265)   
\end{keywords}
\section{Introduction} 
\label{sec_intro}  

The census of baryons at low redshift has revealed that the formation of galaxies and galaxy clusters has been a relatively 
slow process \citep[]{persic92,Fukugita04}. Nearly 90\% of the total number of baryons in the present universe are outside of galaxies,
in circumgalactic gas and in the more  distant intergalactic medium (IGM). A significant fraction ($\sim 30 - 50$\%) of these baryons are 
predicted to have \linebreak temperatures of $T \sim 10^5 - 10^7$~K, resulting from shocks during gravitational collapse leading to structure 
formation \citep[]{Cen99,Dave01}. This warm-hot phase of the intergalactic medium (WHIM) and the associated halo gas is a common outcome of 
structure formation simulations in a $\Lambda$CDM dominated universe. Observations of this diffuse gas phase are crucial for forging a full 
understanding of the physical and ionization conditions of matter in the universe as well for a full accounting of baryons in the present universe.  

{\lya} forest surveys at low-$z$ have shown that the diffuse cool ($T\sim10^4$~K) photoionized gas outside of galaxies contains 
about 30\% of the total baryons \citep[]{Penton04,Lehner07}. Compared to this cooler gas phase, detections of the warm-hot gas have been scarce.
At temperatures of\linebreak $T \geq 10^5$~K, the neutral fraction of hydrogen is expected to be very low $f_{\hi} \equiv N({\HI})/N_{\rm H} \leq 10^{-6}$ 
from collisional ionization \citep[]{Sutherland93}. The warm-hot gas seen in {\lya} will therefore be shallow and thermally broad \citep[$b(\HI) > 40$~{\kms}, 
see e.g.][]{Tripp01}. Searching for these broad-{\lya} absorbers (BLAs) is a proven strategy for probing warm-hot baryons and it has been shown that 
they have a cosmic density of $\Omega_{b}(\rm BLA) \sim 7$\% \citep[]{Richter04,Sembach04,Richter06b,Lehner07}.

An alternate means of detecting the warm-hot gas phase is to search for highly ionized metal line transitions. 
Among the suite of available lines, the \NeVIII\ \lam\lam770, 780~\AA\ doublet is a very sensitive probe of collisionally 
ionized gas at \linebreak $T \geq 10^5$~K. The detection of {\NeVIII}, often associated with {\OVI}, has yielded physical
conditions in several warm-hot absorbers \citep[]{Savage05a,Narayanan09,Savage11,Narayanan11,Tripp11,Narayanan12,Meiring13,Thor13}.
Except for the first \citep[i.e.,][]{Savage05a}, eight of {\NeVIII} detections have come from the high sensitivity high signal-to-noise (SNR)
{\sl Hubble Space Telescope}$(HST)/$Cosmic Origins Spectrograph (COS) spectra in the far-ultraviolet (FUV). In all these studies \NeVIII\ is
shown to be collisionally ionized. However, in an optically thin gas exposed to the extragalactic ionizing background radiation \citep[]{Haardt01}, 
the ionization fraction of \NeVIII\ can become considerable \citep[i.e., $\ge$~0.1, see e.g., Figure~1 of][]{Thor13} under two conditions: 
(a) low-temperature, extremely tenuous photoionized gas (with \linebreak $n_{\rm H} < 10^{-5}$ cm$^{-3}$ and $T < 10^{5}$~K) and (b) warm-hot collisionally 
ionized gas (with $n_{\rm H} > 10^{-5}$ cm$^{-3}$ and $T > 3\times 10^{5}$~K). Therefore, the possibility that the \NeVIII\ originates from 
photoionization cannot be ruled out provided the inferred cloud sizes are not unreasonably large \citep[see e.g.,][]{Savage05a,Narayanan12}.        

Significant progress has been achieved in our understanding of the global properties of low-$z$ {\OVI} absorbers, another
candidate for probing baryons at the lower temperature end of the \linebreak $10^5 - 10^7$~K gas \citep[]{Danforth06,Tripp08,Savage14}. 
From nearly 60 {\OVI} systems currently known at $z < 0.5$, it is estimated that {\OVI} absorbers accounts for $\leq 4 - 10$\% of the total 
baryons, comparable to the baryon fraction trapped within galaxies \citep[]{Tripp00,Savage02,Danforth05,Savage14}. 
The steep power-law column density distribution of the {\OVI} absorbers suggests that the baryon fraction is likely to be more
than doubled if one could detect weak \linebreak ($~N(\OVI) \leqslant 10^{13.0}$~ cm$^{-2}$) {\OVI} systems from higher SNR data. 
There is a lingering concern however on the process that regulates the production of {\OVI} in intervening absorbers. In many cases, 
the {\OVI} and associated {\HI} are consistent with an origin in collisionally ionized warm $T \sim 10^5$~K gas \citep[]{Danforth08}. 
But in a significantly large number of cases  formation of {\OVI} through photoionization cannot
also be ruled out \citep[]{Thom08a,Tripp08,Howk09,Muzahid12,Muzahid14}. Only the former would contribute
towards the accounting of warm-hot baryons since photoionization temperatures are at least an order of
magnitude lower. {\NeVIII} is thought to be free from such ambiguities on the gas phase it traces.    
However, the current sample of {\NeVIII} systems is insufficient. As we wait for the number of {\NeVIII} systems to grow,
it is worthwhile analyzing suitable individual absorption systems to gain insights into the chemical, ionization conditions 
and the physical origin of these absorbers. In that spirit, we report a new detection and analysis of a \NeVIII\ absorber at 
\zabs ~$\sim 0.6$ in the high SNR $HST$/COS spectrum of the quasar PG~1407+265. Previous $HST$ observations of the quasar PG~1407+265 were reported by
\citet{McDowell95} where they discussed the unusual emission line properties of this radio-quiet quasar.
$HST$/FOS observations of the same quasar were reported by \citet{Jannuzi98}.

This article is organized as follows: In Section~\ref{sec_obs} we describe the observation and data 
reduction. Detailed description of the absorbing system and measurements of individual lines are given 
in Section~\ref{sec_measure}. Different possible ionization scenarios are explored in Section~\ref{model}. 
In Section~\ref{sec_summ} we discuss our results and summarize the conclusions. 
Throughout this article, we adopt an $H_0$ = 70~\kms Mpc$^{-1}$, $\Omega_{\rm M}$ = 0.3, and $\Omega_{\Lambda}$ = 0.7 cosmology.
The solar relative abundances for the heavy elements are taken from \citet{Grevesse10}.

\section{Observations and Data reduction}  
\label{sec_obs}  

\begin{figure*} 
\centerline{
\vbox{
\centerline{\hbox{ 
\includegraphics[keepaspectratio=true,scale=0.55,angle=00]{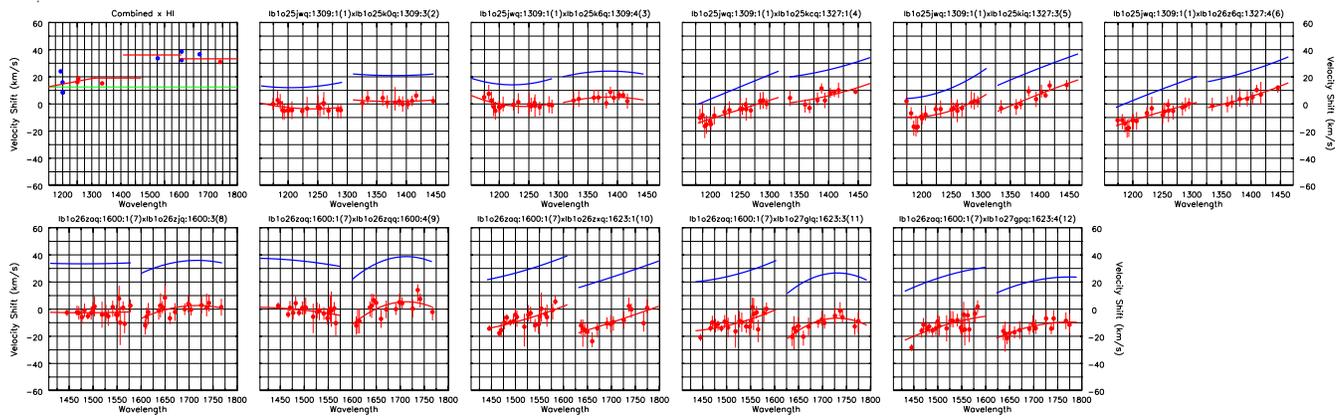}   
}}
}}  
\caption{The first panel shows the offsets between 21-cm data and the
combined, aligned spectra, while the remaining panels give the offsets
needed to bring the spectra into alignment. Red points show the offset
between individual absorption lines in each pair of spectra, as labeled
above each panel. Only the offsets between the reference spectrum
(lb1o04eoq) and the other exposures are shown. The red lines give the
linear fits to these offsets. The blue lines show the combined effect
of relative and absolute (w.r.t 21-cm) offsets, i.e., they show the
offsets that are actually to be applied to each spectrum in order to
align it with the other spectra and with the 21-cm data}          
\label{figa}      
\end{figure*} 

The medium resolution (i.e., $\lambda/\Delta\lambda \sim$~18,000),
high SNR (e.g., $\sim$~40 per resolution element) FUV spectrum of PG~1407+265 (\zem\ = 0.94) was 
obtained using $HST/$COS during observation cycle-17 under program ID: 11741 (PI: Todd Tripp).
The observations consist of G130M and G160M FUV grating integrations covering the wavelength range 1150 -- 1800 \AA. 
The properties of COS and its in-flight operations can be found in \citet{Osterman11} and \citet{Green12}. 
It is now well known that the COS wavelength calibration maybe inaccurate at a level of about $\sim$~20~\kms\ \citep[][]{Savage11, Meiring13}.
For different exposures of the same target, taken at different times and/or with different central wavelength settings, there may be offsets of
up to one resolution element ($\sim 17$~\kms) between corresponding individual spectra. One of us (BPW) has developed an algorithm to correct for this, 
which will be fully described elsewhere \citep[Wakker et al. 2014][in preparation]{}. 
Briefly, this method involves cross-correlating strong
non-black absorption lines produced by the Galactic interstellar medium (ISM) or intergalactic gas between exposures. This gives the relative offsets 
needed to align the spectra at the wavelength of the absorption line. The offsets as a function of wavelength are fitted by a polynomial. 
This fit is used to bring different exposures into alignment before co-addition. Finally, the ISM absorption line velocities in 
the co-added exposures are aligned with the corresponding 21-cm {\HI} emission profile to get a correct absolute wavelength scale.     

In Fig.~\ref{figa} we show the resulting offsets for PG1407+265. 
Clearly, for some combinations of grating central wavelength and fixed-pattern-position (FP-POS) 
settings there are large differential offsets from short to long wavelengths. For example, in panel 4,
the spectrum obtained with grating settings of $\lambda$(cen)=1309~{\AA} and FP-POS=1 is plotted against 
the $\lambda$(cen)=1327~{\AA}, FP-POS=4 exposure. At 1200~\AA, aligning the two spectra requires shifting the $\lambda$(cen) = 1327~{\AA} by $-$10~\kms, 
while at 1300~{\AA} the shift needed is +10~\kms. Applying a constant shift across the spectrum would result in smearing out the lines in the combined 
spectrum. This differential shift is corrected for in the algorithm that we have used.   

The reduced co-added spectra were binned by three pixels, as the COS data in general are 
highly oversampled (six raw pixels per resolution element). All measurements and analysis
presented in this work were performed on the binned data. While binning improves the SNR (per pixel) of the data, 
measurements are found to be fairly independent of binning. Continuum normalization was done by 
fitting the line-free regions with a smooth lower-order polynomial. 

\section{System description and Line Measurements}   
\label{sec_measure} 

\begin{figure*} 
\centerline{
\vbox{
\centerline{\hbox{ 
\includegraphics[height=11.0cm,width=12cm,angle=00]{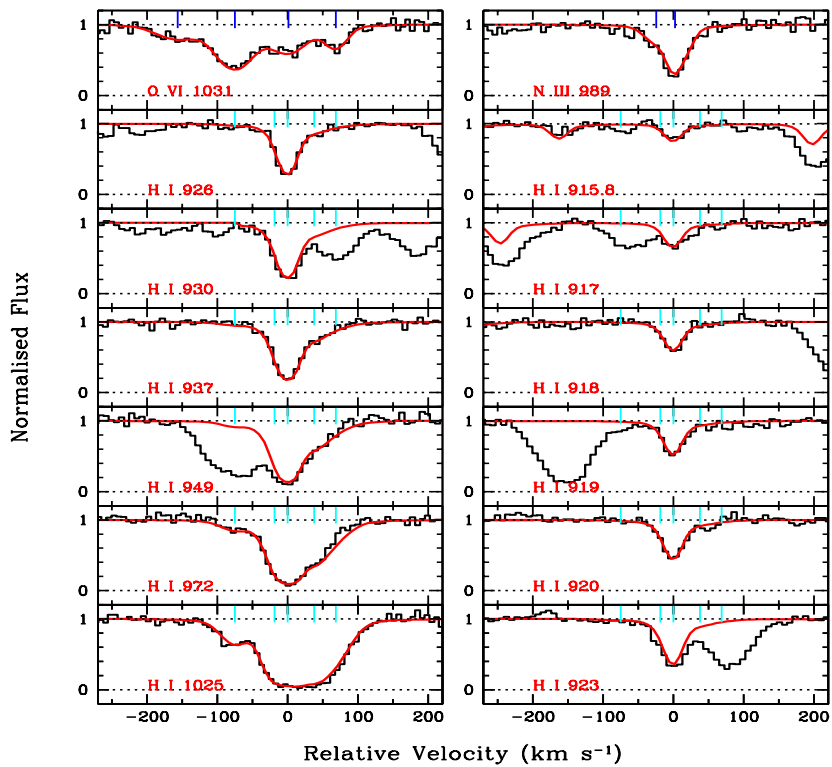}     
}}
}}  
\caption{Continuum-normalized velocity plot for the \HI\ Lyman series lines. Zero 
velocity corresponds to \zabs\ = 0.59961. The best fitting Voigt profiles (smooth 
red curves) are overplotted on top of the data (black histograms). The vertical tick 
marks (in cyan) represent velocity centroids of five \HI\ components used for fitting. 
In the top panels, \OVI\ (in the left) and \NIII\ (in the right) profiles are shown for 
comparison. While \NIII\ closely follows the component structure seen in the core of the \HI\ 
absorption, the \OVI\ profile looks very different suggesting an entirely different origin.    
}              
\label{fig_lys}      
\end{figure*} 

\begin{figure*}
\centerline{
\vbox{
\centerline{\hbox{ 
\includegraphics[height=11cm,width=12cm,angle=00]{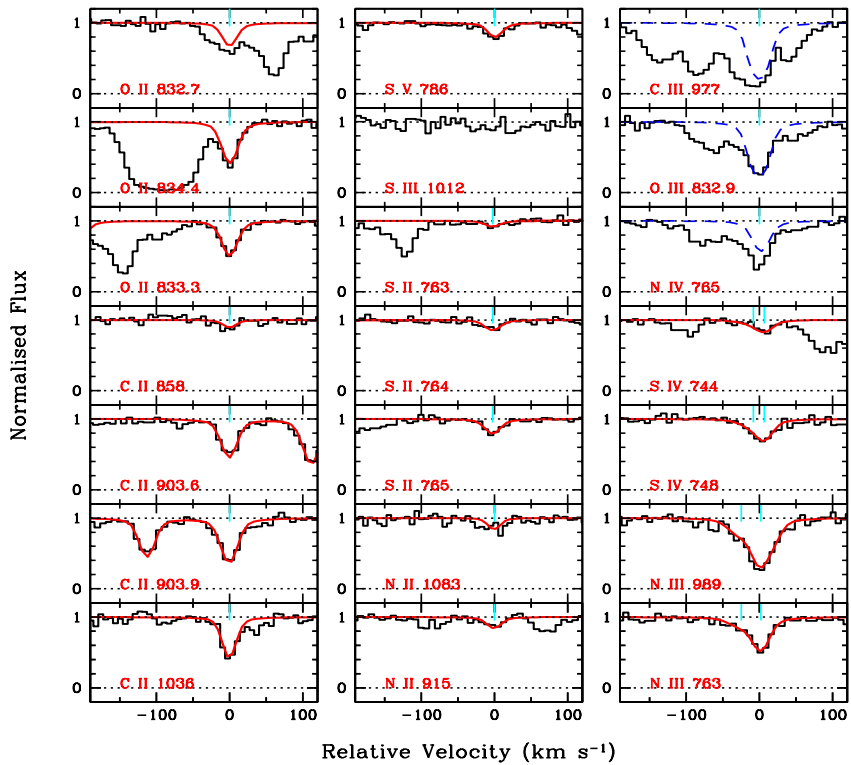}   
}}
}}  
\caption{Continuum-normalized velocity plot for the low-ions. Zero velocity corresponds 
to \zabs = 0.59961. The best fitting Voigt profiles are shown in smooth (red) curves on top 
of the data (black histogram). The vertical tick marks represent velocity centroid for each 
components. \SIII\ is a non-detection. Note that for \NIV, \OIII, and \CIII\ the dashed 
smoothed curves are the synthetic profiles with photoionization model predicted column 
densities (see text).}              
\label{fig_lowion}      
\end{figure*} 

\begin{figure} 
\centerline{
\vbox{
\centerline{\hbox{ 
\includegraphics[height=12cm,width=9.6cm,angle=00]{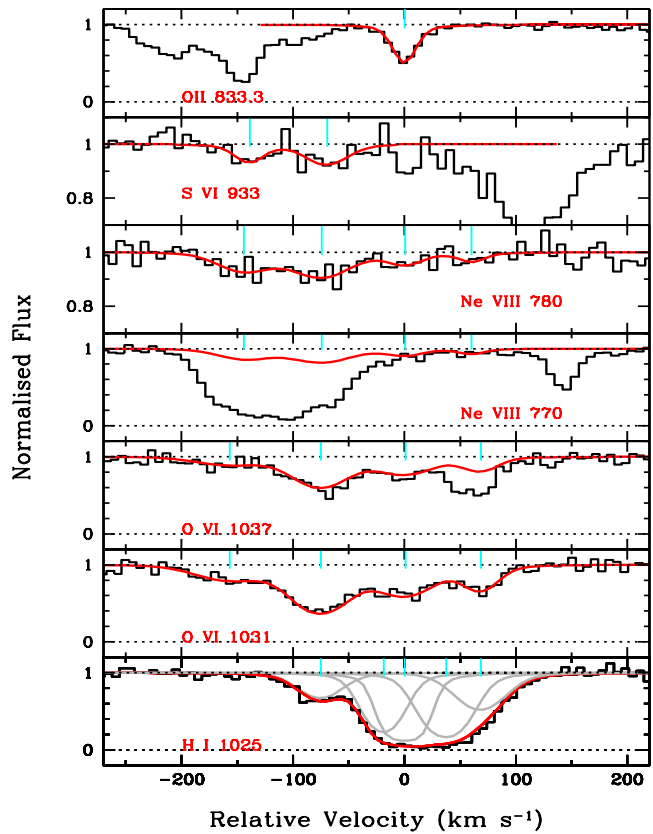}   
}}
}}  
\caption{Same as Fig.~\ref{fig_lowion} but for the high ions. In the bottom panel, 
individual components of \HI~\lam1025 absorption are shown in grey curves. No \lyb\  
absorption is detected corresponding to the high ionization component at $-150$~\kms. 
The other three high ionization components are seen to coincide with \HI. However, \HI\ 
$b$-parameters in these components are narrower compared to \OVI, suggest that they 
do not trace each other. In the top panel the \OII\ profile is plotted to show the difference 
in kinematics of high and low ions.}  
\label{fig_highion}      
\end{figure}

In this work we focus on the system with total $N(\HI) = 10^{16.03\pm0.03}$~cm$^{-2}$ at \zabs = 0.59961 
that also shows \NeVIII\ absorption in the COS spectrum of PG~1407+265. For this system we detect absorption 
from \HI\ Lyman series lines, \OII, \OIII, \OVI, \CII, \CIII, \NII, \NIII, \NIV, \SII, \SIII, \SIV, \SV, and \SVI. 
The metal lines used in our analysis are all detected \linebreak at a $> 3\sigma$ level. Continuum normalized plots of
{\HI} and prominent metal lines are shown in Fig.~\ref{fig_lys}, \ref{fig_lowion}, and \ref{fig_highion}. To extract the
component structure, column densities and Doppler widths we applied Voigt profile fits to the lines using the {\sc vpfit}(v10.0) software package\footnote{http://www.ast.cam.ac.uk/~rfc/vpfit.html}. We adopt the \citet{Kriss11} line spread function (LSF)
to take care of the non-Gaussian nature of the COS LSF. The results of these fits are summarized in Tables~\ref{tab1} and \ref{tab2}.  

The hydrogen absorption associated with this absorber is detected in the Lyman series
lines from {\HI}~\lam 1025 to \lam 915. {\lya} is outside the wavelength coverage of
the COS spectrum. Although \lya\ range is covered in the FOS spectrum available in the $HST$ archive, 
spectral resolution is not high enough to give additional useful information. The higher order Lyman lines are unsaturated, 
providing constraints on the component structure and column densities of {\HI}. A free simultaneous fit to the Lyman 
lines results in a five component kinematic structure listed in Table~\ref{tab1} and shown in Fig.~\ref{fig_lys}. 
The bulk of the neutral hydrogen absorption is at $v_{\rm rel} = 0$~{\kms}, defined with respect to the reference 
redshift \zabs\ = 0.59961, which coincides with the single narrow component absorption seen in the low and intermediate ions. 
Apart from \SVI, \OVI, and \NeVIII, all other ions (having lower ionization energies) detected in this system are taken here to be low/intermediate ions.         

The total $N(\HI)$ measured in this system  is consistent with the absorber being optically thin to
hydrogen ionizing photons. Therefore, the presence of singly ionized species like \CII, \OII, \NII\ etc., 
despite the lack of shielding of ionizing radiation by \HI, makes this system interesting. 
The presence of {\CII} is confirmed by the detection of $\lambda$858,~$\lambda$903.6,~$\lambda$903.9 ~and $\lambda$1036 lines.
The column density and $b$-parameter of {\CII} were estimated by simultaneously free fitting these four lines with a single component.
The ~\OII\ ion transitions are seen at $\lambda$832.7,~$\lambda$833 ~and~ $\lambda$834. However, the \OII\ $\lambda$832.7 line 
is blended and is excluded from profile fitting. 

The ~\NII ~\lam915 and \lam1083 are very weak, but detected at a \linebreak $~\gtrsim 3\sigma$ level. 
The column density and $b$-value for {\NII} were derived by keeping $b$ and $v_{\rm rel}$ of the single component tied 
to the more reliable ~\CII ~fit parameters. Weak absorption from ~\SII ~ is seen through the triplet absorption at ~\lam\lam\lam ~763,~764,~765. 
The line parameters are derived from simultaneous fits to the three lines. 

At the wavelength of \NIII\ 989 we also expect to see ~\SiII ~\lam989, a weaker line of {\SiII}. 
The equivalent width ratio between \NIII\ \lam 989 and \NIII\ \lam 763 is $\sim 2 : 1$. The profiles of the two
lines are also consistent with each other. Using [Si$/$S] solar abundance ratio and measured $\log N(\SII) = 12.53\pm0.03$, the predicted column density
of \SiII\ 989 is $\log N(\SiII) = 11.78\pm0.03$. This column density is too low to produce any detectable absorption. Also, a 3$\sigma$ equivalent width of
\SiII\ 989 is $W = 3$~m\AA.
From this, we infer the contribution from {\SiII} at ~\lam989 to be negligible.

The asymmetry in profile shapes of the two {\NIII} lines is indicative of unresolved multiple components 
contributing to the absorption. A simultaneous formal fit to the \NIII\ \lam989 and \lam763 does reveal two 
components; a weaker component at \vrel $\sim$~ $-$25 ~\kms\ and a stronger component at ~\vrel $\sim$ 2 ~\kms. 
The component at  \vrel $\sim$ 2 ~\kms\ has $z$ and $b$ consistent with those of \CII\ \& \OII\ and contains $\sim$~80\% of the total $N(\NIII)$.

\begin{table} 
\begin{center}  
\caption{Voigt profile fit parameters for \HI\ and low ions.}                   
\begin{tabular}{lrrr}  
\hline 
Species  &     $v_{\rm rel}^{a}$ &  $b$ (\kms)       &   $\log N$ (cm$^{-2}$) \\    
\hline  
\HI      &    $-75\pm0$  &   23$\pm$4   & 14.00$\pm$0.05  \\  
\HI      &    $-19\pm0$  &   18$\pm$2   & 14.61$\pm$0.07  \\  
\HI      &    $  0\pm0$  &   12$\pm$1   & 15.97$\pm$0.02  \\  
\HI      &    $ 37\pm3$  &   22$\pm$6   & 14.77$\pm$0.10  \\  
\HI      &    $ 68\pm0$  &   26$\pm$5   & 14.26$\pm$0.10  \\   
	 &		 &              &                  \\
\CII     &    $  0\pm1$  &    7$\pm$1   & 14.04$\pm$0.07  \\   
\OII     &    $  0\pm1$  &    8$\pm$1   & 14.14$\pm$0.03  \\  
\NII     &    $  0\pm0$  &    7$\pm$0   & 13.16$\pm$0.11  \\  
\NIII    &    $-25\pm16$ &   19$\pm$17  & 13.55$\pm$0.41  \\  
\NIII    &    $  2\pm2$  &   10$\pm$2   & 14.20$\pm$0.09  \\   
\SII     &    $ -3\pm1$  &    9$\pm$2   & 12.53$\pm$0.03  \\  
\SIII    &               &              & $<$13.5         \\ 
\SIV     &    $-9\pm12$  &   19$\pm$0   & 12.70$\pm$0.35  \\          
\SIV     &    $  7\pm3$  &   10$\pm$0   & 13.08$\pm$0.15  \\ 
\SV      &    $ -1\pm1$  &    8$\pm$2   & 12.41$\pm$0.03  \\   
\SiII    &               &              & $<$13.2         \\    
\hline \hline 
\end{tabular} 
\label{tab1}
~ \\ ~\\   
Note -- $^{a}v_{\rm rel}$ is the relative velocity measured w.r.t \zabs\ = 0.59961.             
\end{center}  
\end{table} 

\CIII\ \lam977, \OIII\ \lam832.9 and \NIV\ \lam765 lines are severely contaminated by
absorption unrelated to the system. In Fig.~\ref{fig_lowion} we show the synthetic profiles 
of these lines using the photoionization model predicted column densities (listed in Table~\ref{tab3}) 
and assuming $b(\CIII) = b(\CII)$, $b(\OIII) = b(\OII)$, and $b(\NIV) = b(\NIII)$. The asymmetry in the
\SIV\ \lam744 and \lam748 doublet requires a two component fit, whereas a single component fit was
found adequate for \SV\ \lam786. For the cases of non-detections of \SIII~\lam1012 and \SiII~\lam889 lines, 
we estimate 3$\sigma$ upper limits on column densities from the observed error spectrum assuming a Doppler parameter of $b = $~8 \kms.    

The {\OVI} is detected in both the \lam\lam1031,~1037 lines of the doublet. Compared to the narrow range in velocity ($\Delta v_{90}$ \footnote{see \citet{Muzahid12} for definition.} $\lesssim 90$~{\kms}) over which the low ionization absorption is seen, the {\OVI} is spread over a velocity range of $\Delta v_{90} \sim 270$~{\kms}.
A simultaneous fit to the doublet yields four components at\linebreak ~\vrel = $-$157, $-$75, 0, and 68 \kms\ with respect 
to the reference redshift \zabs\ = 0.59961 (see Fig. \ref{fig_highion}). The $v = 68$~{\kms} component  of \OVI\ \lam1037 is 
blended with absorption unrelated to the system, evident by comparing its strength with the corresponding component in \OVI\ \lam1032 feature. 

The {\NeVIII}~\lam770 line is blended with the {\HI}~\lam949 (\zabs = 0.29741) and {\HI}~\lam1215 (\zabs = 0.01343) absorption. 
The \NeVIII~\lam780 line is however clearly detected with an observed equivalent width of $W_{\rm obs}$ = 52.84$\pm$6.55 m\AA. 
The {\NeVIII} absorption shows similar kinematic spread and velocity structure as {\OVI} (see Fig.~\ref{fig_highion}). 
A free fit to the \NeVIII\ \lam780 feature results in four components with a total column density of \linebreak $\log N$ (\cmsq) = 14.15$\pm$0.18. 
Voigt profile fitting results for the high ions are summarized in Table~\ref{tab2}. It is important to note that the
line centroids of \NeVIII\ components are consistent within 1$\sigma$ of the line centroids of \OVI. This strongly 
suggests that they originate from the same phase of the absorbing gas. A weak \SVI\ \lam933 absorption feature is 
detected at a 3$\sigma$ level (with $W_{\rm obs}$ = 26.35$\pm$7.85 m\AA) in the two highest column density \NeVIII\ 
components. Like \NeVIII\ and \OVI, \SVI\ does not show any resemblance to the component structure of low ions.    

\begin{figure} 
\centerline{
\vbox{
\centerline{\hbox{ 
\includegraphics[height=10.0cm,width=9.0cm,angle=00]{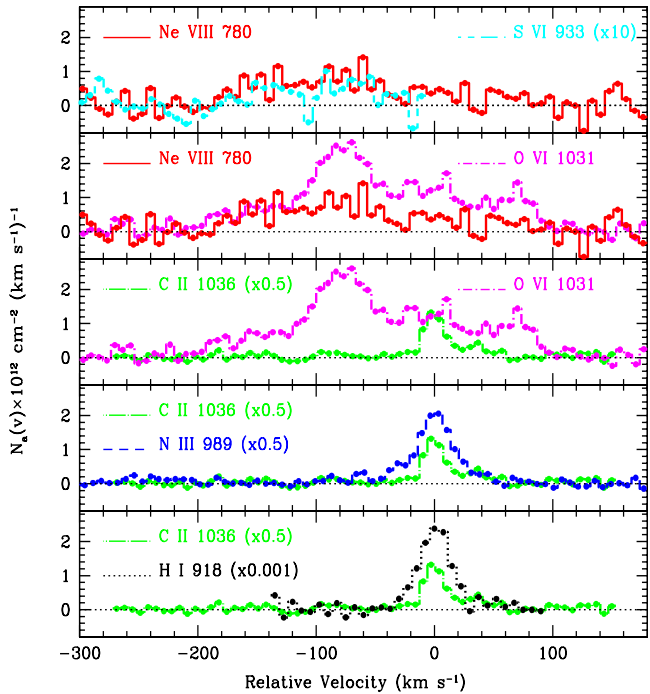}     
}}
}}  
\caption{Apparent column density profiles (in units of 10$^{12}$ cm$^{-2}$ (\kms)$^{-1}$) for 
different lines detected in this system are plotted in velocity scale. Zero velocity 
corresponds to \zabs\ = 0.59961. High and low ionization species show markedly different 
line kinematics.}               
\label{fig_aod}      
\end{figure}

In Fig.~\ref{fig_aod} we show apparent column density profiles \citep[]{Savage91} of several high and low ionization 
absorption lines detected in this system. The profiles of \HI~\lam918, \CII~\lam1036, and \NIII~\lam989 look remarkably
similar and aligned with each other. High ionization species (\SVI, \OVI, and \NeVIII), on the other hand, show a 
completely different profile compared to \CII~\lam1036, but follow each other. This clearly suggests multiphase nature of the absorbing gas.   


\section{Ionization Modelling}
\label{model}

To understand the physical conditions and the chemical enrichment in this absorber, we explore 
different possible ionization scenarios. Under the initial assumption that the gas
is purely photoionized, we generate photoionization models for a range of plausible ionization parameter
defined as $U = n{_\gamma}/n_{\rm H}$, the ratio of ionizing photon number density ($n{_\gamma}$) to the total
hydrogen number density ($n_{\rm H}$). Grids of photoionization (PI) models were generated 
using \linebreak {\CLOUDY} \citep[v13.03, last described by][]{Ferland13} assuming the absorbing 
clouds to be plane-parallel slabs with constant density having solar relative elemental abundances.
The clouds were irradiated with the extragalactic UV background modelled
by \citet[]{Haardt01} at $z$ = 0.6\footnote{An updated version of the extragalactic UV background is presented by \citet{Haardt12}. We
find that our conclusions remain unchanged when we use this updated background radiation.}.
Later on, we also investigate collisional 
ionization equilibrium (CIE) and constant temperature photoionization (PI+CIE) or hybrid models in order to understand the highly
ionized gas phase traced by \SVI, \OVI, and \NeVIII.

\begin{table} 
\begin{center}  
\caption{Voigt profile fit parameters for \NeVIII, \OVI, and \SVI.}                 
\begin{tabular}{lrrr}  
\hline 
Ion           &     $v_{\rm rel}$ (\kms)    &       $b$(\kms)       &   $\log N$ (cm$^{-2}$) \\    
\hline  
\NeVIII       &     $-144\pm 11$    &  29$\pm$16    &   13.65$\pm$0.19  \\  
\OVI\         &     $-157\pm  7$    &  38$\pm$11    &   13.67$\pm$0.09  \\ 
\SVI\         &     $-139\pm  6$    &  14$\pm$10    &   12.39$\pm$0.14  \\  \\  
\NeVIII       &     $ -74\pm  8$    &  31$\pm$15    &   13.80$\pm$0.15  \\  
\OVI\         &     $ -75\pm  2$    &  30$\pm$ 3    &   14.26$\pm$0.03  \\  
\SVI\         &     $ -69\pm  7$    &  22$\pm$10    &   12.58$\pm$0.11  \\ \\  
\NeVIII       &     $   0\pm  9$    &  18$\pm$17    &   13.30$\pm$0.22  \\  
\OVI\         &     $   0\pm  3$    &  31$\pm$ 6    &   13.96$\pm$0.06  \\  \\ 
\NeVIII       &     $  60\pm 10$    &  11$\pm$19    &   13.08$\pm$0.27  \\   
\OVI\         &     $  68\pm  2$    &  18$\pm$ 4    &   13.69$\pm$0.06  \\    
\hline \hline 
\end{tabular} 
\label{tab2}    
\end{center}  
\end{table} 

\subsection{Photoionization model} 
\label{sec:PI}

\begin{figure} 
\centerline{
\vbox{
\centerline{\hbox{ 
\includegraphics[height=9.0cm,width=8.0cm,angle=00]{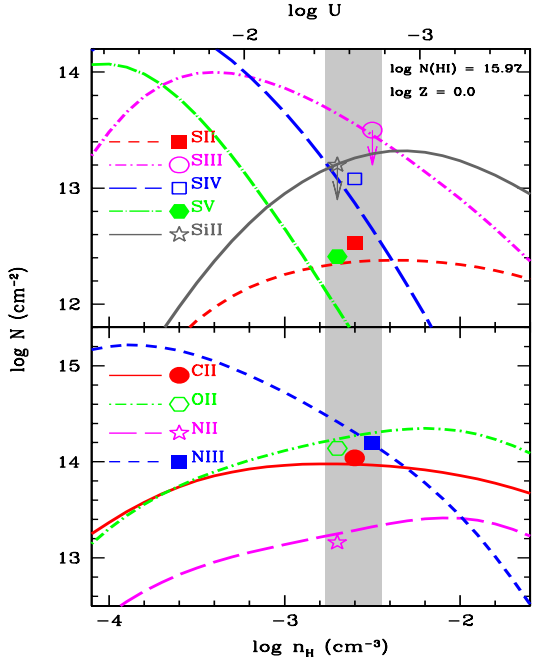}   
}}
}}  
\caption{Results of photoionization model for the low-ionization phase. Different curves 
show the variation of model (with $\log N(\HI)$ = 15.97 and $Z = Z_{\odot}$) predicted 
column densities of different low-ions with density. Corresponding ionization parameter 
is labelled in the top. Measured column densities are indicated by different symbols. 
The shaded region indicates the density range ($-2.7 \leq \log n_{\rm H} \leq -2.5$) for 
which observed column densities are consistent with model predictions.
}           
\label{fig_PImod}      
\end{figure}

The results from photoionization models for $\log~N(\HI) = 15.97$ and solar metallicity are shown in Fig.~\ref{fig_PImod}.
The {\HI} column density is what we measure for the core absorption which coincides in velocity with the single component
absorption in the low ions and also the stronger narrow component in the intermediate ions. 
This component has 88\% of the total $N(\HI)$ measured in this system. 
The best constraint on the ionization parameter comes from $\log [N(\NII)/N(\NIII)] = -1.04\pm0.14$. 
The models are consistent with this observed column density ratio in the interval $-2.7 \leqslant \log U \leqslant -2.5$, which 
corresponds to a density range of \linebreak $n_{\scriptsize\rm H} = (2-3)\times 10^{-3}$~\cmcb\ (the shaded region in Fig.~\ref{fig_PImod}). 
As the gas is optically thin to \HI\ ionizing photons, this density range is not sensitive to the $N(\HI)$ value we assumed in the model. 
The best fit model for low and intermediate ions is obtained for $\log U \sim -2.6$. This corresponds 
to a density of \linebreak $n_{\scriptsize\rm H} = 2.5\times10^{-3}$~\cmcb,
total hydrogen column density of $\log N_{\rm H}$~(cm$^{-2}$) = 18.5, and line-of-sight thickness of $\sim$~410 pc. The density 
we infer here is larger than the inferred $n_{\rm H}$ in the cool (or low ionization) phase of the other known \NeVIII\ absorbers
\citep[see Table 1 of][]{Thor13}. The photoionization equilibrium temperature, as predicted by {\sc cloudy}, is $T\sim10^{3.9}$~K. 
The measured Doppler parameter in the strongest \HI\ component, $b(\HI) = 12\pm1$ \kms\ (see Table~\ref{tab1}), is consistent with this temperature.

In Table~\ref{tab3} we present a comparison of observed and model predicted ionic column densities 
for different species. The model predicted values are calculated at $\log n_{\rm H}=-2.6$, $\log Z/Z_{\odot} = 0$, and $\log N(\HI)$ = 15.97.
Column~4 of the table gives the corrections to the elemental abundances that are required to match the observed column densities. 
It is clear that within 0.2 dex, the relative abundances of C, N, O and S are consistent with a solar abundance pattern. 
Absolute metallicity close to solar with solar relative abundances is consistent with the absorbing gas originating from a 
region having a chemical history similar to the Galactic ISM at the epoch of the formation of the Sun. 
Note that the [N$/$O] = $0.01\pm0.11$ is close to solar in this system.     
It is well known that nitrogen can be produced either by a primary or secondary route depending on
whether the seeds (carbon and oxygen) were produced in the star during the helium burning stage or were already 
present in the star. Primary N is synthesized from intermediate mass stars on the asymptotic giant branch 
(AGB) while secondary N is produced most effectively by all hydrogen burning stars. 
It has been observed that for oxygen abundance [O$/$H]~$\gtrsim-0.4$, the [N$/$O] ratio rises 
with increasing [O$/$H] (secondary N). However at lower metallicities \linebreak (e.g., [O$/$H]~$\lesssim-0.7$), 
the [N$/$O] ratio remain constant at $-0.5$ (primary N). The measured high value of [N$/$O] along with the
solar metallicity suggests that nitrogen in this system is predominantly produced via the secondary route. 
Therefore the absorbing gas was possibly a part of a region that sustained star-formation for a prolonged period.
The chemical enrichment in this absorber is unlike  what is seen in the
high redshift DLAs \citep[]{Petitjean08,Cooke11b,Dutta14} but it 
corresponds to what we see in the interstellar medium (ISM) of our Galaxy  \citep[]{Israelian04, Spite05}.
However no galaxy candidate is detected close to the QSO line of sight in the SDSS image. 
Deeper observations of the PG~1407+265 field are needed to arrive at any firm conclusion.     

\begin{table} 
\begin{center}  
\caption{Observed and model predicted column densities }  
\begin{tabular}{lccccc}  
\hline
\hline
Ion           &     $\log N$ (cm$^{-2}$)   &    $\log N$ (cm$^{-2}$) &   $\rm [X/H]$    &    X   \\ 
	      &	     Observed	           &	Predicted$^{a}$      &                  &        \\
\hline  
\CII          &     $14.04\pm0.07 $    &  13.97     &  	 0.07$\pm$0.07   &   C	\\    
\CIII         &     $-$                &  14.92     &  				\\ 
\CIV          &     $-$                &  13.63     & 				\\    
\OII          &      $ 14.14\pm0.03$   &  14.27     &   $-$0.13$\pm$0.03 &   O	\\  
\OIII         &      $ -$              &  15.16     & 			 &      \\    
\NII          &     $  13.16\pm0.11$    &  13.28     &  $-$0.12$\pm$0.11 &   N	\\     
\NIII         &     $  14.20\pm0.09$    &  14.32     &   		 	\\ 
\NIV          &     $ -$                &  13.22     &  			\\   
\NV	      &     $ -$                &  11.76     &  			\\ 

\SII          &     $  12.53\pm 0.03$   &  12.36     &   0.17$\pm$0.03   &  S 	\\    
\SIII         &     $  <13.5$           &  13.57     & 				\\ 
\SIV          &     $  13.08\pm 0.15$   &  12.91     &  	        	\\ 
\SV           &     $  12.41\pm 0.03$   &  11.76     &                          \\   
\SiII         &     $  <13.2$           &  13.24     &   $< -0.04$	 &  Si  \\  \\    

\SVI          &                         &  11.44     &  \\  
\OVI          &                         &  11.10     &  \\  
\NeVIII       &                         &  $<$10     &  \\      
\hline \hline 
\end{tabular} 
\label{tab3}  
\end{center} 
~\\  
Note -- $^{a}$Photoionization model predicted column densities for \linebreak 
$\log n_{\rm H} = -2.6$, $\log Z/Z_{\odot} = 0$, and $\log N(\HI) =$ 15.97.      
\end{table}

\begin{figure} 
\centerline{
\vbox{
\centerline{\hbox{ 
\includegraphics[height=9.5cm,width=8.5cm,angle=00]{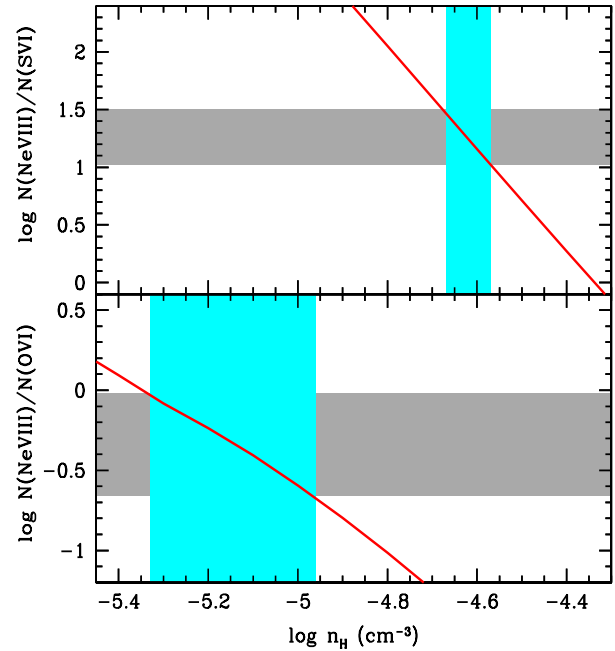}      
}}
}}  
\caption{Results of PI model for high ions. Column density ratios \NeVIII$/$\SVI\ 
(top) and \NeVIII$/$\OVI\ (bottom) are plotted against hydrogen number density.  
The horizontal shaded regions indicate the range of observed column density ratios. 
The vertical shaded regions (in cyan) correspond to the density range over which a PI 
model reproduces the observed ratios. Clearly, \SVI\ is severely underproduced for 
the density range suggested by the \NeVIII$/$\OVI\ ratio.       
} 
\label{fig_PIhot}     
\end{figure} 

\begin{figure} 
\centerline{
\vbox{
\centerline{\hbox{ 
\includegraphics[height=9.5cm,width=8.5cm,angle=00]{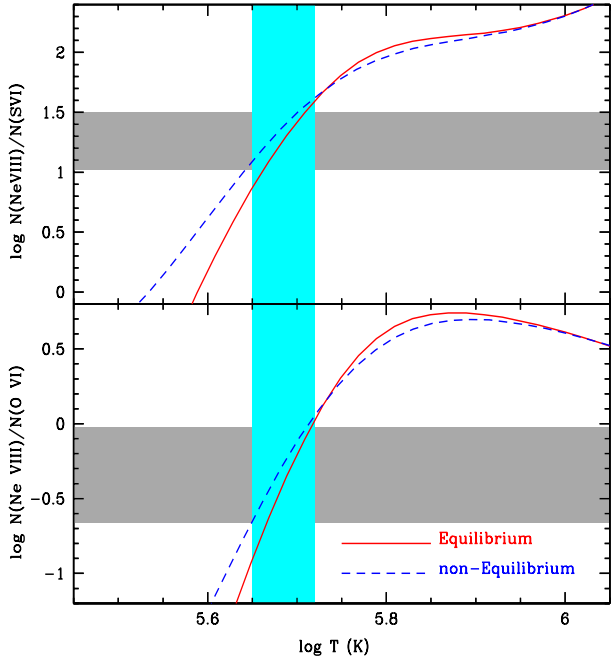}    
}}
}}  
\caption{Column density ratios of \NeVIII$/$\SVI\ (top) and \NeVIII$/$\OVI\ (bottom) 
against gas temperature under CIE (solid curve) and non-CIE (dashed curve) models of 
\citet{Gnat07}. The horizontal shaded region indicates the range in observed column 
density ratios. The vertical shaded region (in cyan) is the range of temperature 
for which CIE model successfully reproduces both the ratios. Non-CIE model suggests 
only slightly wider range in temperature compared to CIE model.
}          
\label{fig_CImod}    
\end{figure} 

It is apparent from Table~\ref{tab3} that, similar to the other known \NeVIII\ absorbers, 
the photoionization model that produces the low ions fails to produce any appreciable column
densities for \SV, \SVI, \OVI, and \NeVIII, even with solar metallicity. Therefore it is obvious
that these highly ionized species originate from a completely different phase of the absorbing gas.
The high ionization phase could be: (a) photoionized with higher ionization 
parameter than the low ionization gas, (b) collisionally ionized with very high 
temperature, or (c) in an environment where both photoionization and collisions due to 
free electrons play an equally important role in deciding the ionization structure and thermal state of the gas. 

We find that a density range of $-5.3 \leqslant \log n_{\rm H} \leqslant -5.0$ (or $-0.2 \leqslant \log U \leqslant 0.1$) is
required to reproduce the observed \NeVIII$/$\OVI\ column density ratios (see bottom panel of Fig.~\ref{fig_PIhot}).
With the help of the ionization simulations, it is fairly straightforward to estimate the line-of-sight thickness for
the high ionization phase given the \NeVIII\ column density:   
\begin{eqnarray}  
& L  =  \frac{N_{\neviii}}{n_{\rm H}f_{\neviii}} \frac{1}{Z} 
        \left(\frac{\rm Ne}{\rm H}\right)_{\odot}^{-1} & \nonumber \\    
    & \simeq 180~{\rm kpc} \left(\frac{N_{\neviii}}{10^{14}}\right) 
    \left(\frac{n_{\rm H}}{10^{-5}}\right)^{-1}  
    \left(\frac{f_{\neviii}}{0.18}\right)^{-1} 
    \left(\frac{1}{Z}\right). &    	 
\label{eqn:size}     
\end{eqnarray}    
Here $f_{\neviii} \equiv n_{\neviii}/n_{\tiny \rm Ne}$ is the 
ionization fraction of \NeVIII, which takes a value of $0.18$ at $\log n_{\rm H}= -5.0$, 
for optically thin conditions. Using Eq.~\ref{eqn:size}, the line-of-sight thickness obtained for the component at $-150$~\kms~ is $\sim$186~kpc assuming solar
metallicity. Such a large size is comparable with 
the sizes of oxygen-rich halos of isolated star forming galaxies at low redshift \citep[$z<0.2$, see][]{Tumlinson11,Werk13,Werk14}.
Evidently, the high ionization gas phase can have a photoionization solution provided the metallicity is high enough \linebreak (i.e., $Z \sim Z_{\odot}$).
Here we argue that the high ionization gas phase cannot have a much lower metallicity (than solar) as we do not see any \lyb\ absorption corresponding to
the high ion component at $\sim -$~150 \kms\ (see Fig.~\ref{fig_highion})\footnote{We obtain a formal 3$\sigma$ upper limit on $N(\HI) <10^{13.4}$~ cm$^{-2}$~ 
using the observed error spectrum.}. The total hydrogen column density associated with this high ionization phase
is $\log N_{\rm H} {\rm (cm^{-2})} = $~19.4, assuming solar metallicity. This is a factor of 10 more than what we have
found for the low ionization phase but is similar to the previously reported values in 
collisionally ionized \NeVIII\ absorbers \citep[see e.g. Table~1 of][]{Thor13}. The \HI\ ionization 
fraction, $f_{\hi} = $~1.2$\times10^{-6}$, as computed by {\sc cloudy}, 
at $\log n_{\rm H}= -5.3$ gives a $N(\HI)$ of $10^{13.3}$~cm$^{-2}$ which is
consistent with the 3$\sigma$ upper limit we estimate for the non-detection of \lyb\ absorption in the $\sim -$~150 \kms\ component.
Our photoionization model solutions are summarized in Table~\ref{tab4}. \\

While our two phase photoionization solutions can explain most of the observed ions (low+high),
none of the phases can reproduce the right amount of $N(\SV)$ and $N(\SVI)$. 
The low ionization phase produces a factor of $\sim$3 less $N(\SV)$
than the observed value. The high ionization phase, on the other hand, produces $\log N < $~11 for both \SV\ and \SVI. 
Fig.~\ref{fig_PIhot} clearly shows that the density range for which observed $\NeVIII/\OVI$ ratio is well 
explained by the model fails to produce enough \SVI. Such a discrepancy can be sorted out by invoking yet another 
phase with density in the range \linebreak $-4.2 < \log n_{\rm H} < -3.5$, where the ionization fractions 
of \SV\ and \SVI\ show their maxima. Note that such a density range will also produce considerable \CIV\ and \NV.
These transitions are not covered with the present COS spectrum. Nonetheless, we note that \CIV\ absorption is clearly detected
in the low resolution FOS spectra \citep[]{Jannuzi98}. \NV, on the contrary, is possibly not detected. 
The measured $N(\CIV)$ and upper limit of $N(\NV)$ are, indeed, consistent with the above quoted density range.
However, because of the low resolution of the FOS spectrum we could not resolve the component structure. Higher resolution data
is certainly required for further analysis of the detailed ionization structure. In the next section 
we investigate if collisional ionization models can successfully explain all the high ions simultaneously in a single phase.

\begin{table*} 
\begin{center}  
\caption{Summary of photoionization model solutions.}        
\begin{tabular}{lccccccc}    
\hline
\hline
Phase&  Species &  $\log n_{\rm H}$  &  $\log U$  &  $\log Z$  &  $\log N_{\rm H}$  &  Size  &  $\log T$ \\     
       &          &  (cm$^{-3}$)   &          & ($Z_{\odot}$) &  (cm$^{-2}$)    &         &    (K)     \\ 
\hline   
Low    & \CII, \CIII, \NII, \NIII, &   $-2.6$($-2.8$)  &  $-2.6$($-2.8$)  &  0.0(0.0)  &  18.5(18.4)   &   410 pc (480 pc)  &  3.9 (4.0)  \\  
       & \NIV, \OII, \OIII, \SiII, &   \\ 
       & \SII, \SIII, \SIV         &   \\ 
High   &  \OVI, \NeVIII            &   $-5.3$($-5.3$)  &  $0.1$($-0.3$)  & 0.0(0.0)   & 19.4(18.8) & 186 kpc(175 kpc) &  4.8 (4.8)  \\  
\hline 
\hline 
\end{tabular}
\label{tab4} 
~\\ ~\\   
Note -- Values in the parenthesis are for the \citet{Haardt12} UV background radiation. \\  
\end{center} 
\end{table*} 

\subsection{Collisional ionization model for the high ions}    
\label{sec:CI}

As discussed by \citet[][see Fig.~21]{Muzahid13} the $N(\NeVIII)/N(\OVI)$ ratios 
found in intervening systems occupy a narrow range that is occupied by the intrinsic absorbers as well.
In addition, they have shown that collisional ionization is indeed a feasible ionization mechanism 
for the intrinsic \NeVIII\ absorbers \citep[see also][]{Muzahid12ne8}. The observed range in the $N(\NeVIII)/N(\OVI)$ ratios 
in the present system is consistent with the earlier measurements in other intervening systems. 
In order to understand whether or not the collisional ionization is a viable process for the highly 
ionized species detected in this system, we consider the models of \citet{Gnat07}. In Fig.~\ref{fig_CImod} we 
show the variation of \NeVIII\ to \OVI\ (and \SVI) column density ratios against the gas temperature under 
equilibrium (CIE) and non-equilibrium (non-CIE) isobaric models. The non-equilibrium processes become
important for high metallicity ($Z > 0.1 Z_{\odot}$) gas. Here we use the isobaric non-CIE model computed at
solar metallicity. However, it is apparent from the figure that CIE and non-CIE models give a fairly 
similar range in temperature \linebreak (i.e., $5.65 < \log T {\rm(K)} < 5.72$) that can explain the
observed $N(\NeVIII)/N(\OVI)$ and $N(\NeVIII)/N(\SVI)$ ratios simultaneously\footnote{We note that non-equilibrium isochoric models also give similar temperature
range.}. As expected, 
this temperature range is consistent with the previous studies of \NeVIII\ absorbers \citep[e.g.,][]{Savage05a,Narayanan11,Narayanan12,Meiring13}.               

As mentioned earlier, we derive a 3$\sigma$ upper limit on \linebreak $N(\HI) < 10^{13.4}$~cm$^{-2}$ for
the high ionization component at $\sim$ $-150$~{\kms}. Using this limit we estimate the metallicity of the 
collisionally ionized hot gas to be\linebreak $\log Z/Z_{\odot} > -1.0$ at $T = 10^{5.7}$~K. Although the lower limit on the metallicity is consistent with that of the cool 
photoionized gas phase, it need not be the case always as it is generally assumed \citep[e.g.,][]{Savage05a,Tripp11,Meiring13}.
The total hydrogen column density 
in this case is \linebreak $ N_{\rm H}<10^{19.6}$ ~cm$^{-2}$. 
The hot phase CIE solution does not 
produce any significant amount of $N(\SV)$, $N(\CIV)$ or any other low$/$intermediate ions, which indicate the presence of an additional phase.

\subsection{Hybrid model for the high ionization gas} 
\label{sec:hybrid}

\begin{figure} 
\centerline{\hbox{ 
\centerline{\vbox{
\includegraphics[height=5.0cm,width=8.4cm,angle=00]{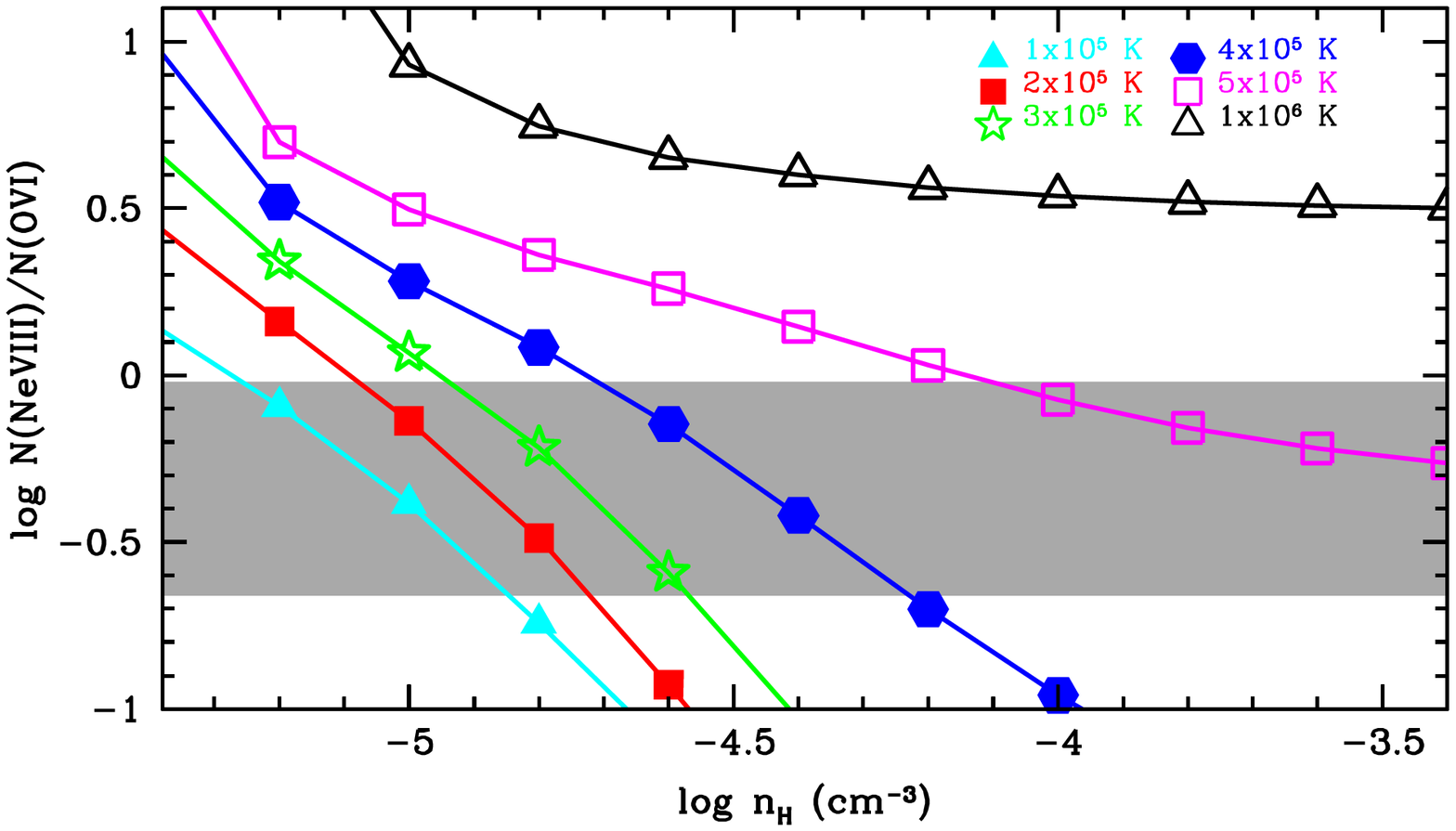}     
\includegraphics[height=5.0cm,width=8.4cm,angle=00]{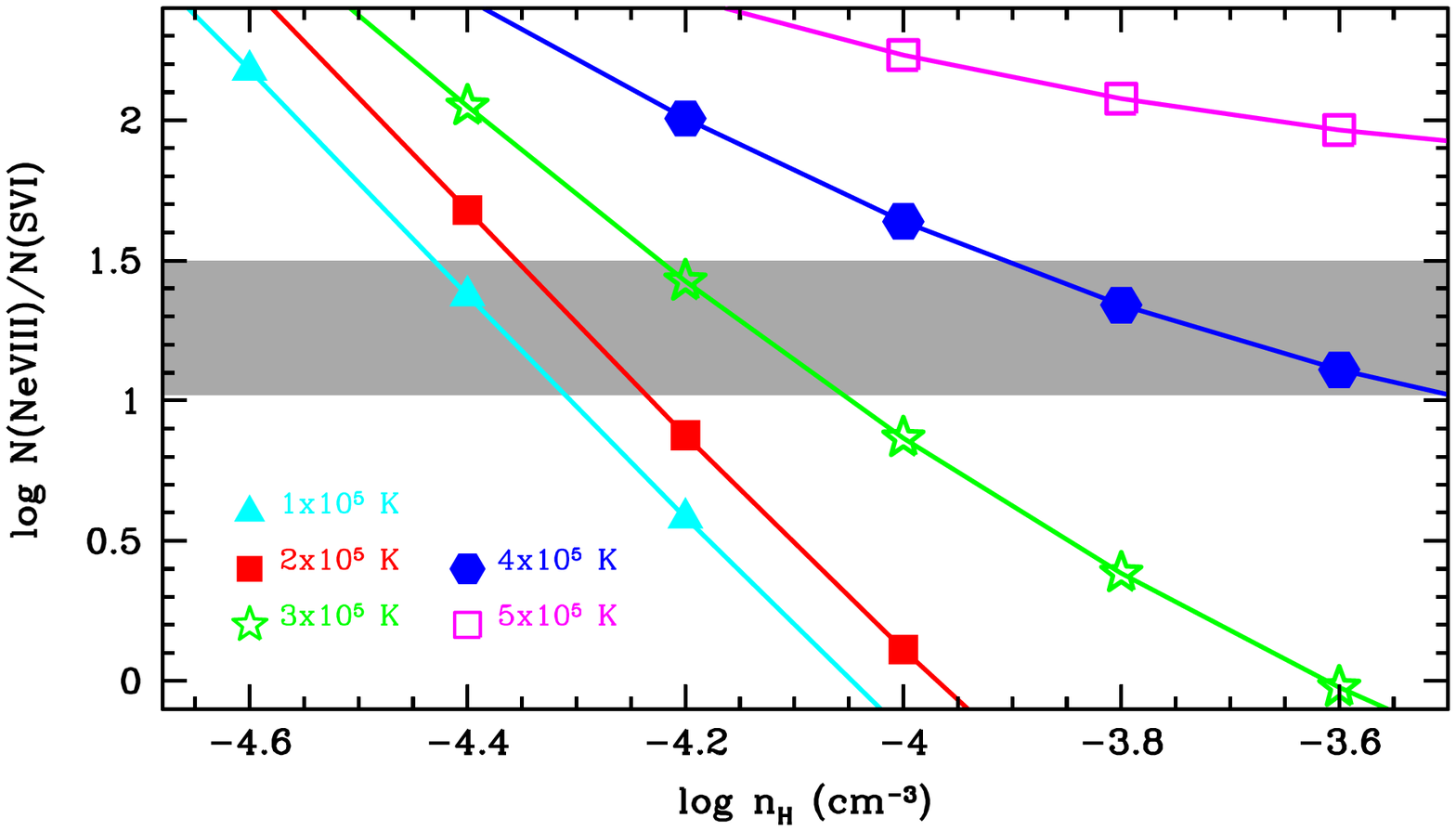}     
}}
}}  
\caption{Column density ratios of \NeVIII$/$\SVI\ (bottom) and \NeVIII$/$\OVI\ (top) 
versus the hydrogen density under hybrid (PI+CIE) models. Different curves correspond to 
different temperatures as indicated in the plot. The horizontal shaded regions show the 
range in observed ionic ratios. 
}             
\label{fig_hybrid}      
\end{figure}

In order to complete our modelling efforts, we explore a hybrid model 
in which we consider a hot gas in the presence of the UV ionizing background. This is a
more realistic scenario than the pure collisional ionization, as the gas cannot be shielded from the 
extra-galactic UV background in this non-Lyman limit absorber. The model grids are computed 
using \CLOUDY\ for an optically thin hot gas at different constant temperatures and exposed 
to the \citet{Haardt01} UV background radiation at $z = 0.6$. With the temperature fixed we try
to understand the contribution of photoionization in a hot gas and how that changes with gas temperature.
In the top panel of Fig.~\ref{fig_hybrid} we show the model predicted \NeVIII $/$\OVI\ column density 
ratio against hydrogen density for different (constant) temperature in the range \linebreak $10^5 -10^6$~K. 
Clearly, the observed ratios cannot be reproduced by the hybrid models for $T> 5\times10^{5}$~K. 
Beyond this temperature collisional ionization dominates over photoionization and $f_{\neviii}$ increases 
rapidly towards its peak. Note that this is the best fit temperature implied by pure collisional ionization
in Section~\ref{sec:CI}. For lower temperatures there are ranges in $n_{\rm H}$ and $T$ that can explain the observed
line ratios. Typically, the observed $N(\NeVIII)$ to $N(\OVI)$ ratios are well reproduced for density $\log n_{\rm H} < -4.2$
and temperature $T < 4\times10^{5}$~K. Models with $T < 4\times10^{5}$~K can also reproduce the observed $N(\NeVIII)/N(\SVI)$, 
however, at much higher density (see the bottom panel of Fig.~\ref{fig_hybrid}). Therefore, like photoionization models, 
a hybrid model also {\sl cannot} reproduce the observed  $N(\SVI)$, $N(\OVI)$, and $N(\NeVIII)$ simultaneously.

\section{Discussion}   
\label{sec_summ}    
  

\subsection{Multiphase Structure}
All the low ions detected in this system show a narrow ($\Delta v_{90}\lesssim$ 90 \kms) and single component absorption profiles 
which are well aligned with the profiles of intermediate ions. Additionally, the line centroid of this 
component matches with that of the strongest \HI\ component. The intermediate ionization species, however, 
show an additional weak component separated by $-25$~\kms, roughly containing 20\% of the observed total 
column densities of these ions. The absorption profiles of the high ionization species are different compared to that of the low ions.
The high ion absorption is comprised of four components 
spread over $\Delta v_{90}\sim$ 270 \kms. The difference in absorption line kinematics between the high and 
low ions clearly indicates the multiphase nature of the absorbing gas. Such a disparity between the 
absorption profiles of high and low ions is in contrast to what has been reported for other \NeVIII\ 
system by \citet[]{Tripp11} and \citet[]{Meiring13}.

\subsection{Low Ionization Gas}
We find that the observed column densities of the low ions and the stronger component of the intermediate 
ions that are aligned with the strongest \HI\ component can be very well explained by a photoionized gas with 
a density of $n_{\rm H}$ = 2.5$\times10^{-3}$ cm$^{-3}$ ($\log U = -2.6$) and an absolute metallicity 
of $Z \sim Z_{\odot}$ irradiated by the UV background radiation contributed by QSOs and galaxies. 
The inferred density range in this system is higher than those derived for the 
similar phase in other known \NeVIII\ absorbers \citep[see e.g., Table~1 of][]{Thor13}. The relatively 
higher hydrogen density with respect to the mean IGM at this redshift (i.e.,   $\bar n_{\rm H} = 7\times10^{-7}$ cm$^{-3}$)
suggests that the gas is possibly associated with a very large over-dense region. The total hydrogen column density associated 
with this cool photoionized phase is   $N_{\rm H} = 10^{18.5}$ ~cm$^{-2}$. This corresponds
to a line-of-sight thickness of $\sim$~410~pc. The model predicted photoionization
equilibrium temperature (i.e., $T\sim10^{3.9}$~K) is consistent with the observed 
narrow doppler parameters of \HI\ and metal absorption lines. This cool photoionized 
phase is unable to explain the observed line strengths of \SV\ and the 
high ionization species: \SVI, \OVI, and \NeVIII.

The relative abundances of C, N, O and S are found to be consistent 
with the solar values within 0.2 dex. Based on [N$/$O] and [O$/$H] ratios we
argue that the nitrogen in this system is predominantly synthesized via secondary production mechanism.
This indicates that the absorbing gas is in close association with a region with a sustained 
star-formation for a prolonged period. Such a chemical enrichment is unlike what is seen in
high redshift DLAs but resembles the Galactic ISM. This, however, is not surprising 
as low-redshift systems are expected to have adequate time for secondary production levels
to begin.  
Recently, using a sample of low redshift ($z < 1$) Lyman limit 
systems (LLS; with 16.2~$\leq \log N(\HI) \leq $~18.5) \citet{Lehner13}
have found that the metallicity distribution of such absorbers exhibit a
bimodal distribution with peaks at $\log Z = -1.6$ and $\log Z = -0.3$ respectively.
The authors have claimed that the metal-rich branch likely traces winds, recycled outflows,
and tidally stripped gas whereas the metal-poor branch is consistent with cold accretion streams 
that fuel gas in star forming galaxies. The system under consideration show a total $\log N(\HI) =$~16.02$\pm$0.03, 
which is close to the lower limit of $N(\HI)$ considered by \citet{Lehner13}. An absolute metallicity of $Z \sim Z_{\odot}$, 
as we estimate for the low ionization phase, indicates that the system belongs to the metal-rich branch and therefore likely to trace an outflowing material.             

The near solar metallicity with solar relative abundances and 
high [N$/$O] ratio found here strongly suggest a connection of this absorber to a star-forming region.
However, no galaxy candidate at the redshift of the absorber is found in SDSS image.
Therefore, this system is not associated with any $L_{\star}$ galaxy, as these will be 
readily visible with the SDSS images. The apparent magnitude limit of SDSS
in the r-band for 95\% completeness is $m_{\rm r}$ $\leqslant$ 22.2\footnote{http://www.sdss3.org/dr10/scope.php/}{http://www.sdss3.org/dr10/scope.php/}. 
This magnitude limit corresponds to a limiting luminosity of $L \sim 0.33 L_{\star}$ at the absorber redshift \citep[]{Montero09}.
Searches for faint galaxies at low impact parameters in deeper images 
will provide further insights on the absorber that shows signature of being associated to star forming region.


\subsection{High Ionization Gas}
A low-temperature (i.e., $T < 10^{5}$~K) optically thin photoionized 
gas with $n_{\rm H} < 10^{-5}$ cm$^{-3}$, exposed to the \citet{Haardt01} UV background at $z \sim 0.5$ can
have considerable $f_{\neviii}$   (e.g., $> 0.1$) \citep[see Fig.~1 of ][]{Thor13}. Here we show that the observed $N(\NeVIII)$ and $N(\OVI)$ 
can be well reproduced by a gas photoionized by the extragalactic UV-background radiation \citep[]{Haardt01}. 
The preferred gas density in such a case is $-5.3 \leq \log n_{\rm H} \leq -5.0$ and the absolute metallicity is 
close to solar (i.e., $Z \sim Z_{\odot}$). From the absence of \lyb\ absorption in one of the high ionization 
component, we argue that the metallicity of the highly ionized gas is indeed near solar. Estimating the \HI\ 
content in different phases (high ionization phase, in particular) of a multiphase system is 
challenging \citep[but see][for a special case]{Muzahid14}. However, \zabs\ = 0.59961 towards PG~1407+265 is a
unique absorption system where we could constrain the $N(\HI)$ associated with the high ionization phase directly from data.          

In this work, we show that an absorber with  $N(\NeVIII) =10^{14}$~ cm$^{-2}$  and with solar metallicity will have a line of
sight thickness of $\sim$ 180 to 200 kpc for the density range suggested by the observed $N(\NeVIII)/N(\OVI)$ ratio in this
system (i.e., $-5.3 \leq \log n_{\rm H} \leq -5.0$). Such sizes are typical of oxygen-rich halos of low redshift star-forming
galaxies \citep[]{Tumlinson11,Werk14}. In fact, the total \OVI\ column density measured in the system 
we presented here is $\log N(\OVI) = 14.57\pm 0.05$, 
which is close to the median  value of $N(\OVI)$ around star-forming galaxies in the sample of \citet{Tumlinson11}.
The total hydrogen column density associated with this phase is $N_{\rm H} \sim 10^{19.4}$~cm$^{-2}$, which 
is very similar to those derived for collisionally ionized \NeVIII\ absorbers \citep[see Table~1 of][for a summary]{Thor13}.    
The photoionization temperature for the high ionization phase is $T\sim 10^{4.8}$~K and is 10 times higher than that 
of the low ionization phase. The density of high ionization gas, on the other hand, is 500 times lower than the low ionization phase. 
Therefore it is difficult to sustain a pressure equilibrium between high and low ionization gas \citep[see also the discussion of][]{Muzahid14}.
Nevertheless, photoionization of \OVI\ and \NeVIII\ is a reasonably good model and we point out that \NeVIII\ {\sl is not an 
unambiguous tracer of collisionally ionized hot gas.} The only trouble with this photoionization model is that it cannot reproduce the \SVI\ 
column density in the same phase. Thus we try to understand these high ions in view of CIE/non-CIE and hybrid models. 
Nonetheless, we cannot rule out the possibility of yet another photoionized phase with density $-4.2 < \log n_{\rm H} < -3.5$.
The \CIV\ and \NV\ absorption features seen in the low resolution FOS spectra are indeed consistent with such a density range.
However, to make any definitive conclusion about this phase it is important to have medium resolution spectrum covering \NV\ and \CIV\ lines.
     

Under CIE, both $\NeVIII/\OVI$ and $\NeVIII/\SVI$ column density ratios are well explained with the temperature 
in the range $5.65 < \log T {\rm (K)} < 5.72$. This narrow range in temperature is consistent with previous
studies where the temperature is derived from $\NeVIII/\OVI$ ratios \citep[e.g.,][]{Savage05a,Narayanan11,Narayanan12}. 
Using the 3$\sigma$ upper limit on $N(\HI)$ as estimated in one of the high ionization component, we derive a total
hydrogen column density, $N_{\rm H} < 10^{19.6}$ cm$^{-2}$ and metallicity, $Z > 0.1 Z_{\odot}$. This clearly
suggests that the metallicity of collisionally ionized warm-hot gas need not be equal to that of cool 
photoionized phase as it is sometimes assumed \citep[e.g.,][]{Tripp11,Meiring13}. Note that the collisionally 
ionized gas does not produce any considerable amount of $N(\SV)$. This is in contrast to what have been found 
by \citet{Tripp11} for the \zabs\ = 0.927 system towards PG~1206+459 where there was no \OVI\ coverage. 
Next, we found that a hybrid (PI+CIE) model can explain the observed $N(\NeVIII)/N(\OVI)$ ratio for a range of
densities and temperatures, but cannot explain observed line strength of \SVI\ in the same phase. Therefore, for the present system, we do not 
find a compelling case to support the CIE instead of PI models as both of them require similar number of distinct ionization phase.

\subsection{Sizes of \NeVIII\ absorbers}
Several previous studies of \NeVIII\ systems \citep[see e.g.,][]{Savage05a,Narayanan11,Narayanan12} have found that the required 
line-of-sight thickness of the absorber became unreasonably large (i.e., $>$ few Mpc) in order to reproduce the observed $N(\NeVIII)$ via
photoionization by the extra-galactic UV background. The \NeVIII\ system at \zabs = 0.20701 towards the QSO HE 0226-4110 analyzed by \citet{Savage05a}
has a $N(\NeVIII)$ to $N(\OVI)$ ratio very similar to the present system. Assuming $\log Z$ = $-0.5$ and using old version of the 
extra-galactic UV background radiation \citep[i.e.,][``HM96"]{Haardt96}, the author inferred a line-of-sight thickness of $\sim$ 11 Mpc. The 
line-of-sight thickness become a factor of 10 less (i.e.,~$\sim$ 1.3 Mpc) when we use the latest extra-galactic UV background \citep[i.e.,][``HM12"]{Haardt12}
and $\log Z$ = $-0.5$. Note that the metallicity of this absorber is not well constrained from data but was assumed to be same as measured in the 
moderately ionized phase. We estimate a line-of-sight thickness of $\sim$~420~kpc assuming a solar metallicity.
The line-of-sight thickness 
of the \zabs = 0.32566 system towards the QSO 3C 263 becomes $\sim$ 700 kpc (instead of $\gtrsim$ 1 Mpc as inferred by \citet{Narayanan12}) when we use ``HM12" 
extra-galactic UV background. 
Therefore, we conclude that the line-of-sight thickness of the previous \NeVIII\ absorbers, irradiated with latest UV 
background radiation (``HM12'') get significantly reduced, but still, is very large and hence a photoionized origin can be ruled out.

 \section{Summary}
 We present a detailed analysis of a \NeVIII\ system at\zabs~ = 0.59961 in the high SNR COS spectrum of the QSO PG~1407+265 (\zem~= 0.94). 
In addition to \NeVIII, absorption lines from low (\CII, \NII, \OII\ and \SII), intermediate (\CIII, \NIII, \NIV, \OIII, \SIV\ and \SV), and high 
(\SVI, \OVI\ and \NeVIII) ions are detected at greater than $3\sigma$ significance. 
 The \NeVIII~\lam780 line is detected at 
 a 8$\sigma$ level with an observed equivalent width of $W_{\rm obs} = $ 52.8$\pm$6.5 m\AA. 
All these metal ions with significantly
different ionization potentials provide stringent constraints on the ionization condition and/or thermal state of the gas. Moreover, several
unsaturated higher order \HI\ Lyman series lines (\HI \lam937 to \lam915) provide an accurate estimate of the neutral hydrogen content and 
hence of the metallicity of the absorber. We briefly summarize the important results:

 \begin{itemize}

  \item[1.]The velocity profiles of the low (\CII, \NII, \OII\ and \SII) ions are narrow and are perfectly aligned with the 
  intermediate (\CIII, \NIII, \NIV, \OIII\ and \SIV) ions. Photoionization model solutions show that the low and intermediate ions origin
   in a metal enrinched ($Z \sim Z_{\odot}$) over-dense gas of density $n_{\rm H}$ = 2.5$\times10^{-3}$~\cmcb\ having a compact size $\sim$ 410~pc. 
   The density measured in this system is larger than the inferred $n_{\rm H}$ in the low ionization phases of the other known \NeVIII\ absorbers.
    The estimated relative abundances of C, N, O and S are consistent with the solar values within 0.2 dex
 indicating that the photoionized gas was part of a region that sustained star-formation for a prolonged period.

  \item[2.]The highly ionized species \NeVIII\ and \OVI\ show a very different spread out velocity profiles compared to the low ions.
  A diffuse ($\sim$~180 kpc), low density ($-5.3 \leqslant \log n_{\rm H} \leqslant -5.0$), photoionized gas with solar metallicity
  can reproduce the observed column densities of  \NeVIII\ and \OVI.
  The above low and high photoionized gas cannot explain the observed column densities of \SV, \SVI\ and \CIV\ (detected in the FOS spectrum), which, require
   an intermediate photoionized gas phase with density $-4.2 < \log n_{\rm H} < -3.5$.

  \item[3.]A single phase collisional ionization model with $5.65< \log T <5.72$ can reproduce the observed column densities of \SVI, \OVI, 
 and \NeVIII\ simultaneously. This temperature range is consistent with all the previous \NeVIII\ absorbers. The metallicity of the collisionally
 ionized hot gas is $Z > 0.1 Z_{\odot}$. This hot phase also fails to reproduce \SV\ and$/$or \CIV\ column densities which indicate the presence of 
 an intermediate phase. Therefore, whether one uses CIE or PI models, the present system requires at least three distinct phases to reproduce
 the observed column densities.

  \item[4.]The line-of-sight thickness of the \NeVIII\ absorbing gas, irradiated with latest updated extra-galactic 
  UV background radiation \citet{Haardt12} 
   is reasonable so the CIE models cannot be favoured simply based on the inferred line-of-sight thickness of the absorbing gas.
   This is unlike what have been found for all the other known \NeVIII\ absorbers. 
  Using solar metallicity and latest UV background radiation we checked the sizes of the previously known 
  \NeVIII\ absorbers under photoionization. The sizes of the absorbers get reduced (compared to the sizes reported using Haardt \& Madau (1996) UV 
  background) but still are very large and hence we rule out the 
  possibility of a photoionized origin for these \NeVIII\ absorbers.

  \item[5.]The high chemical enrichment of the absorber strongly connects it to a star-forming region. However, no galaxy 
 candidate is detected in the vicinity of the QSO line-of-sight in the SDSS image. Deep search for faint dwarf galaxies with low impact parameters will provide
 further insights on the absorber-galaxy connections.
 \end{itemize}

\section{Acknowledgement}
This work has made use of $HST$/COS data and therfore we are grateful to all the people associated with the design and construction of COS onboard $HST$.
TH acknowledges IUCAA for travel support, library 
and computational facilities during the period of this work.
AP acknowledges IUCAA visiting associateship program and a DST-SERB grant.

\def\aj{AJ}%
\def\actaa{Acta Astron.}%
\def\araa{ARA\&A}%
\def\apj{ApJ}%
\def\apjl{ApJ}%
\def\apjs{ApJS}%
\def\ao{Appl.~Opt.}%
\def\apss{Ap\&SS}%
\def\aap{A\&A}%
\def\aapr{A\&A~Rev.}%
\def\aaps{A\&AS}%
\def\azh{AZh}%
\def\baas{BAAS}%
\def\bac{Bull. astr. Inst. Czechosl.}%
\def\caa{Chinese Astron. Astrophys.}%
\def\cjaa{Chinese J. Astron. Astrophys.}%
\def\icarus{Icarus}%
\def\jcap{J. Cosmology Astropart. Phys.}%
\def\jrasc{JRASC}%
\def\mnras{MNRAS}%
\def\memras{MmRAS}%
\def\na{New A}%
\def\nar{New A Rev.}%
\def\pasa{PASA}%
\def\pra{Phys.~Rev.~A}%
\def\prb{Phys.~Rev.~B}%
\def\prc{Phys.~Rev.~C}%
\def\prd{Phys.~Rev.~D}%
\def\pre{Phys.~Rev.~E}%
\def\prl{Phys.~Rev.~Lett.}%
\def\pasp{PASP}%
\def\pasj{PASJ}%
\def\qjras{QJRAS}%
\def\rmxaa{Rev. Mexicana Astron. Astrofis.}%
\def\skytel{S\&T}%
\def\solphys{Sol.~Phys.}%
\def\sovast{Soviet~Ast.}%
\def\ssr{Space~Sci.~Rev.}%
\def\zap{ZAp}%
\def\nat{Nature}%
\def\iaucirc{IAU~Circ.}%
\def\aplett{Astrophys.~Lett.}%
\def\apspr{Astrophys.~Space~Phys.~Res.}%
\def\bain{Bull.~Astron.~Inst.~Netherlands}%
\def\fcp{Fund.~Cosmic~Phys.}%
\def\gca{Geochim.~Cosmochim.~Acta}%
\def\grl{Geophys.~Res.~Lett.}%
\def\jcp{J.~Chem.~Phys.}%
\def\jgr{J.~Geophys.~Res.}%
\def\jqsrt{J.~Quant.~Spec.~Radiat.~Transf.}%
\def\memsai{Mem.~Soc.~Astron.~Italiana}%
\def\nphysa{Nucl.~Phys.~A}%
\def\physrep{Phys.~Rep.}%
\def\physscr{Phys.~Scr}%
\def\planss{Planet.~Space~Sci.}%
\def\procspie{Proc.~SPIE}%
\let\astap=\aap
\let\apjlett=\apjl
\let\apjsupp=\apjs
\let\applopt=\ao
\bibliographystyle{mn}
\bibliography{mybib}
\end{document}